# Biophysical effects and neuromodulatory dose of transcranial ultrasonic stimulation.

*Tulika Nandi (1,2), Benjamin R. Kop (1), Kasra Naftchi-Ardebili (3), Charlotte J. Stagg (4,5), Kim Butts Pauly (3)[a], Lennart Verhagen (1)[a]*

   a. *Joint last authors*

   1. *Donders Institute for Brain Cognition and Behaviour, Radboud University, Nijmegen, The Netherlands*
   2. *Department of Human Movement Sciences, Vrije Universiteit Amsterdam, Amsterdam, The Netherlands*
   3. *Department of Radiology, Stanford University, Stanford, CA, USA*
   4. *Wellcome Centre for Integrative Neuroimaging, FMRIB, Nuffield Department of Clinical Neurosciences, University of Oxford, Oxford, UK*
   5. *Medical Research Council Brain Network Dynamics Unit, Nuffield Department of Clinical Neurosciences, University of Oxford, Oxford, UK*

## Abstract

*Transcranial ultrasonic stimulation (TUS) has the potential to usher in a new era for human neuroscience by allowing spatially precise and high-resolution non-invasive targeting of both deep and superficial brain regions. Currently, fundamental research on the mechanisms of interaction between ultrasound and neural tissues is progressing in parallel with application-focused research. However, a major hurdle in the wider use of TUS is the selection of optimal parameters to enable safe and effective neuromodulation in humans. In this paper, we will discuss the major factors that determine both the safety and efficacy of TUS. We will discuss the thermal and mechanical biophysical effects of ultrasound, which underlie its biological effects, in the context of their relationships with tunable parameters. Based on this knowledge of biophysical effects, and drawing on concepts from radiotherapy, we propose a framework for conceptualising TUS dose.*

*Introduction*

Transcranial ultrasonic stimulation (TUS) is a cutting-edge non-invasive brain stimulation technique with much higher spatial resolution and deep brain stimulation capability compared to electromagnetic techniques. Ultrasound has a variety of biophysical effects on tissues, which in turn drive neurobiological mechanisms that lead to neuromodulation. Although the direct links between biophysical effects and neuromodulation are only partly elucidated, fundamental research focused on filling these gaps is progressing in parallel with application-focused research. The major challenge of how to most appropriately set the large number of tunable TUS parameters, recently defined in the ITRUSST standardised reporting consensus[1], remains to be overcome. For fundamental research, manipulating these parameters may allow the development and testing of hypotheses concerning underlying biophysical effects. In application-focused research, establishing a framework for conceptualising 'dose' based on adjustable parameters and knowledge of their associations with biophysical effects is essential for designing studies and ensuring reproducibility. Here, we summarise the relationships between tunable TUS parameters and the thermal and mechanical biophysical effects of ultrasound. Subsequently, we propose a framework for conceptualising dose and a definition based on tunable parameters.

*Biophysical effects of ultrasound and their relationships with tunable parameters*

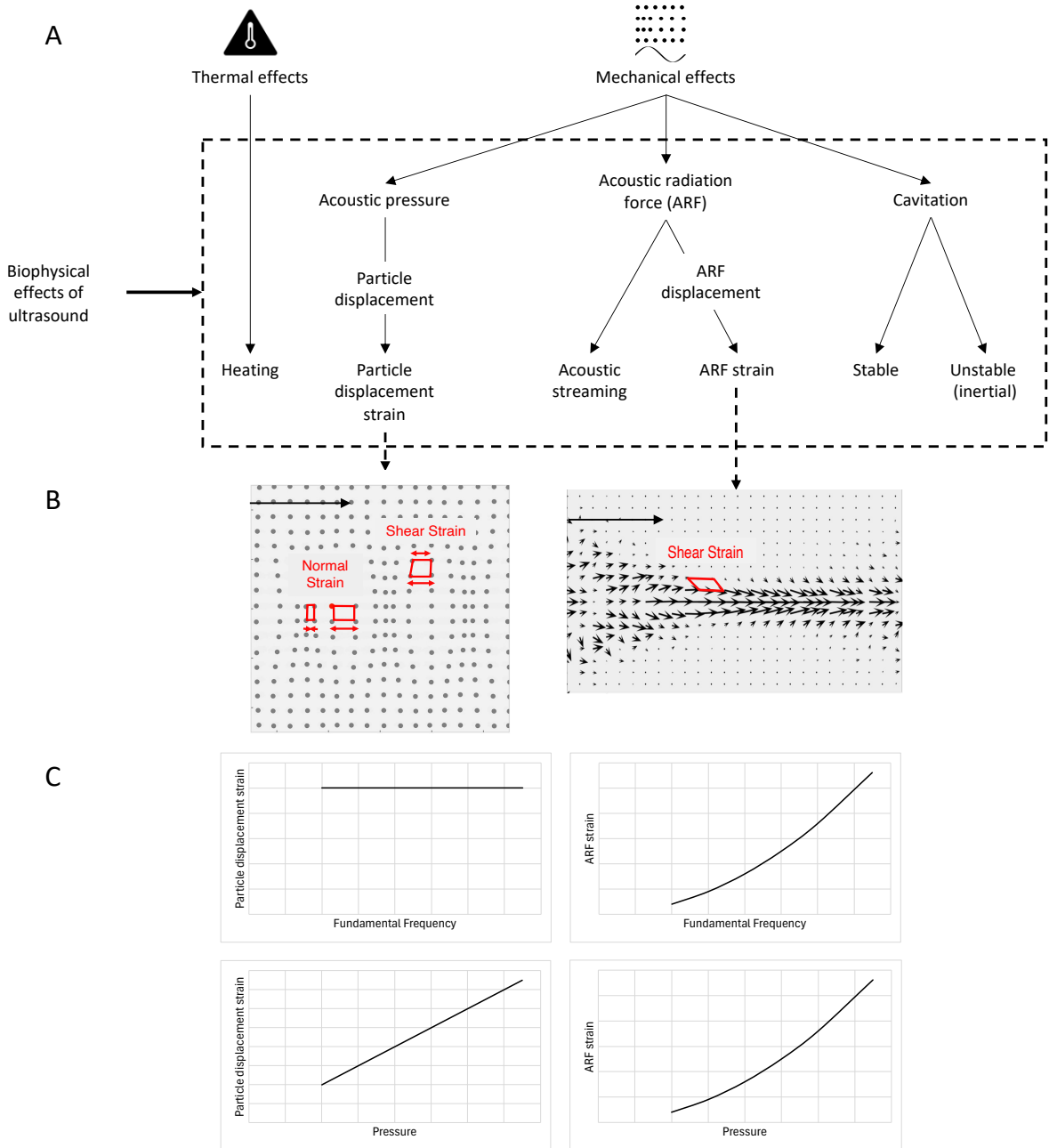

Figure 1: A) Biophysical effects of ultrasound, including thermal (left) and mechanical (right) mechanisms. B) Depictions of normal and shear strain caused by particle displacement (left) and acoustic radiation force (right). Black arrows show the direction of the ultrasound beam. Red arrows show the expansion/compression of the tissue. C) Theoretical relationships between particle displacement (left) and ARF (right) strains and fundamental frequency (top) and pressure (bottom).

Ultrasound can interact with tissues, including neurons and glia, through thermal and mechanical mechanisms outlined in Figure 1A. It is likely that ultrasound affects the underlying brain tissue via multiple mechanisms simultaneously. The relative contributions of these mechanisms, however, can be influenced by adjusting the stimulation parameters.

The thermal effects, i.e. heat or energy deposition, are proportional to the integral of ultrasound intensity over time. Heat or energy deposition can be increased by increasing the intensity, the pulse duration, or both. If heat deposition were the only mechanism that affected the temperature, then any two sets of pulses with the same energy deposition would have the same temperature rise. However, heat is continuously removed, via conduction to adjacent areas and blood circulation, albeit at a relatively slow rate on the order of seconds. This means that when a pulse or a pulse train is extended to seconds or longer, the temperature rise is lower than for the same energy deposition over a much smaller time period. Heating is known to have neuromodulatory effects[2,3], but for transcranial in vivo applications, in which skull heating is a limiting factor, protocols are specifically designed and chosen to limit heating.

Mechanical effects of TUS can be divided into particle displacement strain, acoustic radiation force (ARF) strain, acoustic streaming, and cavitation. Strain refers to the physical deformation of neurons or parts thereof (e.g., the cell membrane), relative to their original configuration when they experience a pressure or force. Particle displacement strain is caused by the acoustic pressure: tissue is stretched and compressed as the pressure wave passes by. The deformation parallel to the direction of the beam results in normal strain and dominates over shear strain which is caused by unequal deformations at different locations on the axis perpendicular to the beam (Figure 1B). Particle displacement is directly proportional to the applied acoustic pressure and inversely proportional to the fundamental frequency ($f_0$). Therefore, particle displacement is larger at lower frequencies (Figure 1C). However, at lower frequencies, this displacement is also occurring over a larger wavelength. As a result of these opposing relationships, particle displacement strain is constant across frequencies (Figure 1C).

ARF strain, on the other hand, is caused by the acoustic radiation force which is a force along the ultrasound beam in the direction of propagation, exerted on absorbing or reflecting tissues in the US path[4]. In contrast to particle displacement strain, ARF strain is predominantly shear strain (Figure 1B). When estimated using some simplifying assumptions[4], the ARF strain is directly proportional to the intensity (i.e., pressure squared) at the focus (Figure 1C). Using the same logic as above, we estimate strain by normalising the displacement to the wavelength and see that while the ARF itself is proportional to $f_0$, the ARF-induced strain is proportional to $f_0$ squared (Figure 1C).

There are two other key points at play: the temporal and spatial response of these two strains. The particle displacement strain varies temporally with each cycle of the ultrasound pulse. In contrast, while the US is on, the ARF is temporally constant at a given point within the medium[5], with the displacement rising exponentially to a maximum over several milliseconds, and decaying exponentially over several milliseconds. Pulsing the US leads to fluctuations in the ARF at the pulse repetition frequency (PRF), a much lower frequency than the fundamental frequency. With respect to the physical location, particle displacement strain is highest at the focus. In contrast, ARF strain is highest where the ARF displacement changes most rapidly, typically in areas adjacent to the focus (Fig B).

In an attenuating medium such as neural tissue, a spatial gradient of ARF is observed as US is absorbed along the beam path. When US passes through a fluid medium, this ARF gradient can cause bulk flow of fluid known as acoustic streaming. The velocity of streaming, and therefore any biological effects caused by it, are highly dependent on the viscosity of the medium, and also the physical boundaries and constraints on the medium.

Cavitation refers to both the pulling of dissolved gas out of sonicated tissues to form gas bubbles, and the oscillation of gas bubbles, including the aforementioned emerging bubbles, existing gas bubbles, or injected microbubbles. Inertial cavitation, a special case of cavitation in which bubbles grow in size and eventually collapse leading to large temperature rises and sudden release of energy, can cause tissue damage and is intentionally avoided during neuromodulation applications. Inertial cavitation is a threshold event i.e., it occurs when a specific threshold, defined by a combination of bubble size, $f_0$ and peak negative pressure, is exceeded. The pressure threshold for inertial cavitation is lower at lower $f_0$[6].

***Conceptualising dose***

We can build on our current knowledge of ultrasound biophysics and parameters to propose hypotheses about the biophysical effects underlying ultrasonic neuromodulation[7], and empirically test their relationships[8]. The development of integrative theoretical frameworks and comprehensive empirical studies is an area of active research, and no consensus has yet been reached. As noted earlier, in practice, efforts to elucidate the underlying mechanisms will proceed in parallel with efforts to optimise TUS effects for basic science and clinical applications. Therefore, we propose a theoretical framework for conceptualising dose, that is agnostic of the underlying biophysical effect.

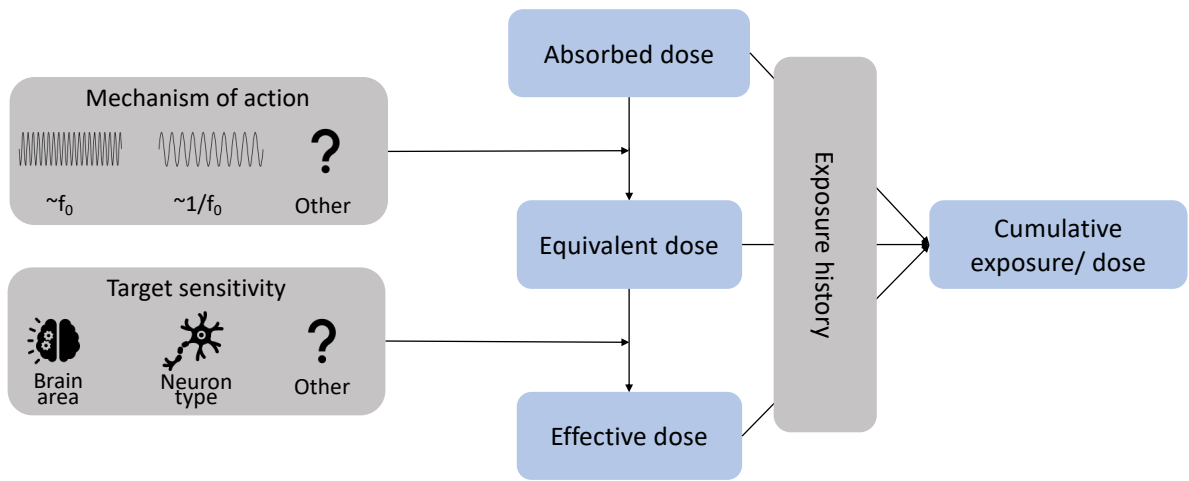

Figure 2: Theoretical framework for conceptualising transcranial ultrasonic stimulation dose.

Knowledge of dose-response relationships is crucial for designing research protocols and for clinical applications. Initial attempts to define acoustic dose have drawn on ideas from radiotherapy[9], and in line with radiotherapy, a distinction must be made between exposure and dose. Shaw et al.[9] describe exposure as the 'energy flux or the acoustic pressure of an ultrasonic wave incident on the region of interest'. Exposure is determined by the source transducer and acoustic properties of the medium. The derated or simulated acoustic pressures/intensities reported in neuromodulation studies are, in essence, measures of exposure. In other words, this is the pressure that the target region of interest is exposed to, after taking into account any attenuation caused by tissues between the transducer and target.

Dose, on the other hand, depends on the interaction of the incident ultrasound wave with the neural tissues of interest. In radiotherapy, there are further distinctions between the absorbed, equivalent, and effective dose. Absorbed dose refers to the amount of energy deposited in the tissues. When a region of interest is exposed to ultrasound, only a fraction of the energy carried by the ultrasound wave is absorbed within this region, while the rest passes through. Absorbed dose is a useful quantity when determining the thermal effects of ultrasound since heating is proportional to the absorbed fraction. An analogue for mechanical effects is that ARF at a given spatial location relates to the energy absorbed at that location.

In radiotherapy, equivalent dose additionally accounts for the differing effects of different types of radiation. The analogue for ultrasound stimulation applications is the biophysical effect underlying any neuromodulatory effects. For instance, for thermal effects, equivalent dose, and therefore temperature rise, simply increases with increasing intensity. However, if neuromodulation is primarily driven by mechanical effects, given a constant exposure, the equivalent dose would increase with increasing $f_0$ for ARF strain-dependent effects, but be independent of $f_0$ for particle displacement strain-dependent effects.

Effective dose accounts for the differing sensitivities of various target tissues, in this case, the target neurons or brain region, to the same equivalent dose. In the context of thermal effects, the same temperature rise may have different effects in different brain regions. If mechanical effects are mediated by mechanosensitive ion channels, given the same equivalent dose, the effective dose required to elicit similar responses would differ based on the density of such channels in different target regions[10]. Another example is the variation in stiffness between brain regions[11], and due to ageing and pathology[12–14]. Given the same acoustic pressures and ARFs, the strain and consequent biological effects will vary based on the stiffness of the target tissue. Importantly, the concept of 'effective dose' should consider the multiple levels of organisation at which ultrasound exerts effects, from biophysics to cellular biomechanisms, to circuit-level neurophysiology, to the human brain and behaviour, and to clinical outcomes (for review, please see[15]). Ultrasound can have different, even opposing, effects on individual levels. For example, a specific ultrasound protocol and dose might selectively facilitate inhibitory neurons, leading to neural excitation but circuit-level inhibition[16]. Effective dose is not a property of the stimulation parameters alone but of the interaction with neural systems. Indeed, investigations of effective dose will require an integrative approach bridging across all levels, from biophysics to cellular biomechanisms, to circuit-level neurophysiology, to the human brain and behaviour, and to clinical outcomes.

Finally, when ultrasound leads to neuroplastic effects, the effect of ultrasound will depend on the history of exposure or dosing, and therefore, it is important to record cumulative dose. For TUS, we therefore propose a framework that includes absorbed, equivalent, effective, and cumulative dose (Figure 2). While this framework is useful to clarify and further our understanding of ultrasonic neuromodulation, there is currently, limited consensus about the biophysical mechanisms and neuronal or regional sensitivities of ultrasound. As our knowledge of TUS expands, we expect the definitions of dose to evolve. For instance, equivalent dose might be weighted by fundamental frequency while effective dose might be weighted by intrinsic mechanosensitive ion channel density.

We propose the integral of amplitude over time as a preliminary definition for dose. Until we have reached an informed consensus on the mechanisms underlying ultrasonic neuromodulation, we cannot distinguish between exposure and dose and thus we can use them interchangeably as 'exposed dose'. All other parameters being constant, increasing pressure increases heating, ARF, particle displacement, and the probability of cavitation. Therefore, the intensity over time is likely to be an important factor in dose, irrespective of the underlying biophysical effects. Time is included in the definition because prolonged exposure could either lead to cumulative effects on a single neuron, for instance by allowing greater time for ion movement and changes in membrane potential[17], or increase the probability of recruiting additional neurons.

Existing empirical data support our proposed definition. However, several aspects of dose have yet to be comprehensively explored. While both in vitro and in vivo data suggest a scaling of response with intensity and duration, the exact nature of the dose-response relationship is currently unclear. Specifically, dose-response relationships may not always be linear, not even monotonic[18]. Further, the identification of thresholds and ceilings is crucial to avoid underdosing and minimise side effects. Lastly, it is important to consider the potential impact of dose-rate, over and above total dose[19]. Note, for example, that the same integral of intensity over time can be achieved by applying either a low-pressure wave for a prolonged duration or a high-pressure wave for a short time-period. However, for thermal effects, greater temperature increases are achieved when energy is delivered over a short time period (a high dose-rate), compared to a low dose-rate where more time is available for the heat to dissipate. In terms of mechanical effects, the dose-rate might interact with the viscoelastic14 properties of neurons. Indeed, the impact of dose-rate should be a key focus area for future empirical research. Given the complexities of the above factors, we advise against conflating dose with pulsing regimes, such as intending to change dose, by changing the pulse duty cycle. For example, given a constant amplitude, a change from 20% to 40% duty cycle could be considered a doubling of dose, while a change from 20% to 100% duty cycle would fundamentally alter the pulsing regime to a continuous wave which may even be less effective[15,18]. Accounting for these considerations, our proposed definition of dose is a pragmatic and neutral starting point, while our understanding of ultrasonic neuromodulatory mechanisms continues to improve.

*Conclusions*

In this paper, we summarise how the bioeffects of ultrasound can be tuned by adjusting the parameters of the application. We provide a theoretical framework for conceptualising dose and propose a preliminary definition for US dose that is agnostic to the underlying biophysical effect.

**Funding:** KN is supported by a grant from the National Institutes of Health (NIH R01 EB032743). KBP is supported by grants from the National Institutes of Health (NIH R01 MH131684, NIH R01NS112152, NIH R01 EB032743). CJS holds a Senior Research Fellowship, funded by the Wellcome Trust (224430/Z/21/Z). LV is supported by a VIDI fellowship (18919) funded by the Dutch Research Council (NWO), by an Open Call HHT (HiTMaT-38H3) funded by Holland High Tech, and is a co-applicant on an EIC Pathfinder project (CITRUS, 101071008) funded by the European Innovation Council (EIC) and on an ERC Advanced project (MediCoDe) funded by the European Research Council (ERC).

**Declarations of interest:** KBP has no competing interests related to the reported work. Unrelated to the reported work, KBP has a relationship with MR Instruments, having received equipment on loan, and with Attune Neurosciences as a consultant. LV has no competing interests related to the reported work. Unrelated to the reported work, LV has a relationship with Brainbox Initiative as a member of the scientific committee, with Nudge LLC, having received consulting fees, with Sonic Concepts Ltd, having received equipment on loan, and with Image Guided Therapy, having received equipment on loan.